# Non perturbative determination of the running coupling constant in quenched SU(2).

G.M. de Divitiis, R. Frezzotti, M. Guagnelli and R. Petronzio

Dipartimento di Fisica, Università di Roma *Tor Vergata*

and

INFN, Sezione di Roma II

Viale della Ricerca Scientifica, 00133 Roma, Italy

July 4, 1994

**Abstract**

Through a finite size renormalization group technique we calculate the running coupling constant for quenched SU(2) with a few percent error over a range of energy varying by a factor thirty. The definition is based on ratio of correlations of Polyakov loops with twisted boundary conditions. The extrapolation to the continuum limit is governed by corrections due to lattice artifacts which are proportional to the square of the lattice spacing and appears rather smooth.

# 1 Introduction

Lattice gauge theories can challenge the determination of the strong coupling constant obtained from high energy experiments. The most precise data are from the LEP collaborations who derive a value for $\alpha_s$ with a six percent error at the $Z$ energy scale. This error turns into a fourty percent error on $\Lambda_{QCD}$ because asymptotic freedom makes low energy values of the coupling constant very sensitive to their high energy values. Precise tests of the standard model require a higher precision for $\alpha_s$, of the order of one percent at the $Z$ mass. Indeed, global fits to LEP data show a variation of the strong coupling constant with the top mass of the order of one to two percent for a twenty GeV variation of the heavy quark mass. In this paper we present a lattice gauge theory calculation of the running coupling constant in quenched SU(2) which shows that the required precision can indeed be reached. The method follows the work of refs. [1, 2, 3, 4] where a finite size scaling technique was first applied to the estimate of the running coupling constant. Our work differs from the ones quoted above for the choice of the observable which defines the renormalized coupling. The definition of refs. [2, 3, 4] is based on the Schrödinger functional while ours is related to the ratio of correlations of Polyakov loops with twisted boundary conditions. The choice of twisted boundary conditions was dictated by the necessity of performing a weak coupling (small $g^2$) lattice perturbative calculation to determine the relation between our definition and the other ones. Indeed, according to the definition, the calculation must be performed on a finite size lattice where twisted boundary conditions eliminate the problem of torons. This calculation was completed and the results reported in ref. [7]. Section 2 contains the generalities of the definition and reviews the method. Section 3 discusses the details of the simulation and the use of lattices with an odd number of points in spite of the fact that our definition refers to correlations of Polyakov loops taken at a distance of half the lattice size. Section 4 contains the results for the running coupling constant including the extrapolation to the continuum limit and the calibration of the energy scale. A first successful comparison of these results with those of the Schrödinger functional is made. A detailed comparison will be made elsewhere [8]. Finally, we present the results for the $\beta$–function.

## 2 The definition and the method

The calculation adopts the following boundary conditions:

$$\begin{aligned} U_\mu(r+\hat{\nu}L) &= \Omega_\nu U_\mu(r)\Omega_\nu^\dagger \qquad && \hat{\nu}=\hat{x},\hat{y} \\ U_\mu(r+\hat{\nu}L) &= U_\mu(r) && \hat{\nu}=\hat{z},\hat{t} \end{aligned} \tag{1}$$

with $U_\mu(r+\hat{\nu}L)$ a "replica–link", *i.e.* a link in a site displaced by the lattice size in the $\nu$ direction and $\Omega_x$, $\Omega_y$ two unitary matrices satisfying appropriate commutation rules (see refs. [5, 6] for details). We have imposed twisted boundary conditions on two directions which eliminate the finite volume degeneracy of toron configurations with the perturbative vacuum and allow a standard perturbative expansion [7]. Polyakov loops which wrap around the lattice in the twisted directions (*e.g.* the $x$–direction) are defined by:

$$P_x(y,z,t) = Tr([\prod_{j=0}^{L-1} U_x(x=j,y,z,t)]\Omega_x e^{i\frac{\pi}{L}y}) \tag{2}$$

where $\Omega_x$ is the twisting matrix in the direction of the loop. This form is determined by the requirement of gauge and translation invariance. The above definition produces correlations of "twisted" loops which start at order $\alpha_s$ in the weak coupling limit. In addition, the disconnected term in the correlation is identically zero. However, such a correlation cannot be taken straightly as a definition of the coupling because of linear divergences, characteristic of Wilson lines, which spoil the continuum limit. A proper definition is obtained by dividing such a correlation by a similar one between Polyakov loops extended in a direction with periodic boundary conditions ($\hat{z}$ or $\hat{t}$).

A crucial ingredient of the method is the definition of the energy scale of the renormalized coupling related to the total lattice size, which allows a finite size scaling recursive scheme. The distance between Polyakov loops should be taken equal to some fixed fraction of the lattice size, half in our case. By defining the coefficient of the tree level result extrapolated to the finite–volume continuum limit as $T(\infty)$, the definition for $g^2(aL)$ reads:

$$g^2(aL) = \frac{1}{T(\infty)} \frac{\langle \sum_{y,z} P_x(0,0,0) P_x^\dagger(y,z,t=L/2)\rangle}{\langle \sum_{x,y} P_z(0,0,0) P_z^\dagger(x,y,t=L/2)\rangle} \tag{3}$$

where the correlation in the denominator has to be understood as a full one, disconnected part included.

The ratio of the $\Lambda$ parameter of this definition ($\Lambda_{TP}$) to the $\overline{MS}$ one, calculated in ref. [7], is given by:

$$\Lambda_{TP}/\Lambda_{\overline{MS}} = 1.6136(2) \tag{4}$$

By replacing the normalization with the tree level result $T(\infty)$ at $L = \infty$ in eq.(3) with the tree level $T(L)$ at finite $L$, one obtains a definition which differs by terms of order $a^2$ and therefore is characterized by the same $\Lambda_{TP}$.

The recursive scheme, already explained at length in the original papers, will be only summarized here:

- on a volume with $L$ points per side, and physical linear size $L_{phys} = La$ at a value of $\beta = 4/g_0^2$ large enough to be close to the continuum limit, one calculates the renormalized coupling $g^2(La)$;
- by repeating the same calculation for the case with $2L$ points, one gets the coupling $g^2(2La)$ renormalized at a length scale which is twice larger;
- the previous two steps are repeated for larger values of $\beta$, *i.e.* for smaller values of the lattice spacing, while the total initial physical volume $L^4_{phys}$ is kept constant. This is achieved by changing the number of lattice points and the value of the bare coupling so that the value of the renormalized coupling stays constant;
- the set of calculations at different lattice spacings are extrapolated to zero lattice spacing giving the value of the change of the renormalized coupling by a change in the length scale by a factor 2 in the continuum;
- the continuum value of the renormalized coupling at the larger scale is taken as the input value for a new step and the procedure is iterated until one reaches a physical volume which is large enough to make the calculation of a physical quantity, which will calibrate the value of the lattice spacing, free from finite size effects;
- the normalization of the lattice spacing on the largest volume sets the scale of the running coupling constant in physical units.

## 3 The details

The simulation was made by using an action with ordinary periodic boundary conditions and a tower of frustrated plaquettes in the $x$–$y$ plane, which can be shown [6] to be equivalent to a standard action with the twisted boundary conditions defined in eq.(1).

The updating scheme (HFKP) was the modification of the standard Creutz's heat bath algorithm introduced in refs. [9, 10] which improuves the acceptance of the original algorithm at high values of $\beta$. In order to fight the critical slowing down at large volumes and high $\beta$, the HFKP algorithm was followed by a number of overrelaxed iterations increasing with the volume up to 12 for the case $L = 16$ at $\beta = 3.08$ . The Polyakov loops measurement is affected by large fluctuations: the number of measurements, the auto–correlation time $\tau$ and the relative error $\epsilon$ obtained for different values of $\beta$ and different volumes are given in table 1. In order to reduce the fluctuations, we have used the "one link integral" technique by replacing the product of links in eq. (2) with the product of the values of the links averaged in the "heat bath" of their neighbours. For the SU(2) group one obtains the following analytic expression:

$$\langle U_\mu(x)\rangle_{OLI} = \frac{kI_0(k) - 2I_1(k)}{kI_1(k)}\overline{V}^\dagger_\mu(x) \tag{5}$$

where $k = \beta \det(V_\mu(x))^{1/2}$, $\overline{V}_\mu(x) = V_\mu(x)/\det(V_\mu(x))^{1/2}$, $V_\mu(x)$ is the "staple" relative to the link $U_\mu(x)$ and $I_n$ is the modified Bessel function of order $n$. The variance of the correlation of the twisted Polyakov loops improuves with the "one link integral" technique by a factor which increases from 2 to 3 for $\beta$ decreasing from 4 to 3.

In spite of the definition which involves the time distance $L/2$, we did also use lattices with an odd number of points. Consider our definition of $g^2(aL)$ (eq. (3)) calculated at time distance $(L-1)/2$ in lattice units. Because of periodic boundary conditions in the time direction such a definition is actually the arithmetic average of the one at distance $t = (L-1)/2$ and of the one at distance $t = (L+1)/2$ and therefore differs from the one at $t = L/2$ by an even function of $1/L$. This guarantees that it will have the same $\Lambda$ and then the same continuum limit. However, in order to plot the results of odd lattices on the same extrapolation curve of even lattices, we need a definition which reproduces the one at $t = L/2$ up to terms at least

| $\beta$ | $L$ | $N_{ov}$ | $N_{meas}$ $(K)$ | $\tau$ | $\epsilon$ |
|---|---|---|---|---|---|
| 2.8350 | 4 | 5 | 1272 | 4 | 0.0014 |
| 2.9800 | 6 | 6 | 1912 | 7 | 0.0016 |
| 3.0800 | 8 | 8 | 1992 | 10 | 0.0018 |
| 2.8349 | 8 | 8 | 240 | 10 | 0.0064 |
| 2.9103 | 10 | 8 | 160 | 15 | 0.0085 |
| 2.9745 | 12 | 10 | 280 | 19 | 0.0071 |
| 3.0305 | 14 | 12 | 160 | 25 | 0.0094 |
| 3.0800 | 16 | 12 | 112 | 28 | 0.0126 |
| 3.3946 | 4 | 4 | 952 | 4 | 0.0016 |
| 3.5200 | 6 | 4 | 952 | 5 | 0.0020 |
| 3.6200 | 8 | 6 | 952 | 8 | 0.0023 |
| 3.3946 | 8 | 6 | 560 | 10 | 0.0034 |
| 3.4563 | 10 | 7 | 240 | 13 | 0.0055 |
| 3.5179 | 12 | 8 | 560 | 20 | 0.0064 |
| 3.6200 | 16 | 10 | 240 | 36 | 0.0106 |
| 3.9650 | 4 | 4 | 150 | 3 | 0.0034 |
| 4.0300 | 5 | 4 | 1024 | 4 | 0.0014 |
| 4.0900 | 6 | 4 | 637 | 4 | 0.0019 |
| 4.1800 | 8 | 4 | 156 | 5 | 0.0048 |
| 3.9762 | 8 | 6 | 240 | 6 | 0.0043 |
| 4.0300 | 10 | 6 | 304 | 8 | 0.0042 |
| 4.0873 | 12 | 6 | 304 | 13 | 0.0054 |
| 4.1780 | 16 | 8 | 144 | 19 | 0.0087 |

Table 1: The number of over–relaxation sweeps *per* heat-bath sweeps $N_{ov}$, the number of measurements $N_{meas}$, the auto–correlation times $\tau$ and the relative errors $\epsilon$ obtained for different values of $\beta$ and $L$

of order $1/L^4$. This is possible if our definition admits a Taylor expansion around $t = L/2$. Both numerical and perturbative investigations show that this is the case. A linear combination can then be formed of operators at $t = (L-1)/2$ and $t = (L-3)/2$ which has the desired property that its expectation value differs from the one of the operator at $t = L/2$ by terms of order $1/L^4$ only. If we rewrite the definition at $t = L/2$ (eq. 3) in the form

$$g^2(aL) = \frac{1}{T(\infty)} \frac{f_x(\beta, L, L/2)}{f_z(\beta, L, L/2)} \quad , \tag{6}$$

where the expectation values $f_x$ and $f_z$, which depend on the bare coupling constant $\beta$, the lattice size $L$ and the correlation time distance $L/2$, are defined by comparison with eq. (3), the required linear combination is obtained by making in eq. (6) the following replacements:

$$\begin{aligned} f_x(\beta, L, \frac{L}{2}) &\rightarrow \frac{9}{8} f_x(\beta, L, \frac{L-1}{2}) - \frac{1}{8} f_x(\beta, L, \frac{L-3}{2}) \\ f_z(\beta, L, \frac{L}{2}) &\rightarrow \frac{9}{8} f_z(\beta, L, \frac{L-1}{2}) - \frac{1}{8} f_z(\beta, L, \frac{L-3}{2}) \end{aligned} \tag{7}$$

We will see that points obtained with this definition do align with those for even lattices in the curves showing the extrapolation to zero lattice spacing limit and indirectly confirm the smallness of order $1/L^4$ corrections.

## 4 The extrapolation to the continuum limit

Getting rid of the lattice cutoff is the most delicate aspect of the calculation. Table 2 contains the results at fixed renormalized coupling for the value of the coupling renormalized at a double length scale, for different values of the lattice spacing, *i.e.* on a fixed physical volume for different values of $L$. The same results are reported in figure 1. The extrapolation to the continuum limit appears rather smooth. We outline the points obtained on odd lattices which align with the even lattice points. The value of the coupling at the scale $L_{phys}$ should be kept constant in the extrapolation procedure, but this is possible only within the statistical errors. In order to obtain the value of $\beta$ which corresponds to the reference value of $g^2(L_{phys})$ we have made an interpolation among the $\beta$ values nearby and the resulting error is taken as the uncertainty on the value of the reference coupling. By using a two loop expression for the $\beta$–function, the errors on the coupling at scale $L_{phys}$ are converted into errors on the coupling at the scale $2L_{phys}$ and are added to its statistical fluctuations as additional and independent statistical fluctuations: in table 2 the errors on the coupling at the scale $2L_{phys}$ include these two contributions.

| $\beta$ | $L$ | $g^2(La)$ | $g^2(2La)$ |
|---|---|---|---|
| 3.9762 | 4 | 2.075 | 2.334(12) |
| 4.0300 | 5 | 2.075 | 2.381(11) |
| 4.0873 | 6 | 2.075 | 2.410(15) |
| 4.1780 | 8 | 2.075 | 2.417(22) |
| 3.7109 | 4 | 2.430 | 2.848(10) |
| 3.7620 | 5 | 2.430 | 2.942(18) |
| 3.8210 | 6 | 2.430 | 2.942(14) |
| 3.9200 | 8 | 2.430 | 2.975(16) |
| 3.3946 | 4 | 3.068 | 3.809(14) |
| 3.4563 | 5 | 3.068 | 3.853(23) |
| 3.5179 | 6 | 3.068 | 3.892(19) |
| 3.6200 | 8 | 3.068 | 3.975(34) |
| 3.0879 | 4 | 4.108 | 5.527(22) |
| 3.1607 | 5 | 4.108 | 5.552(37) |
| 3.2235 | 6 | 4.108 | 5.719(27) |
| 3.2770 | 7 | 4.108 | 5.602(49) |
| 3.3300 | 8 | 4.108 | 5.621(48) |
| 2.8349 | 4 | 5.597 | 8.065(42) |
| 2.9103 | 5 | 5.597 | 8.126(61) |
| 2.9745 | 6 | 5.597 | 8.161(48) |
| 3.0305 | 7 | 5.597 | 8.207(69) |
| 3.0800 | 8 | 5.597 | 8.039(87) |

Table 2: The renormalized couplings for different $\beta$ and $L$

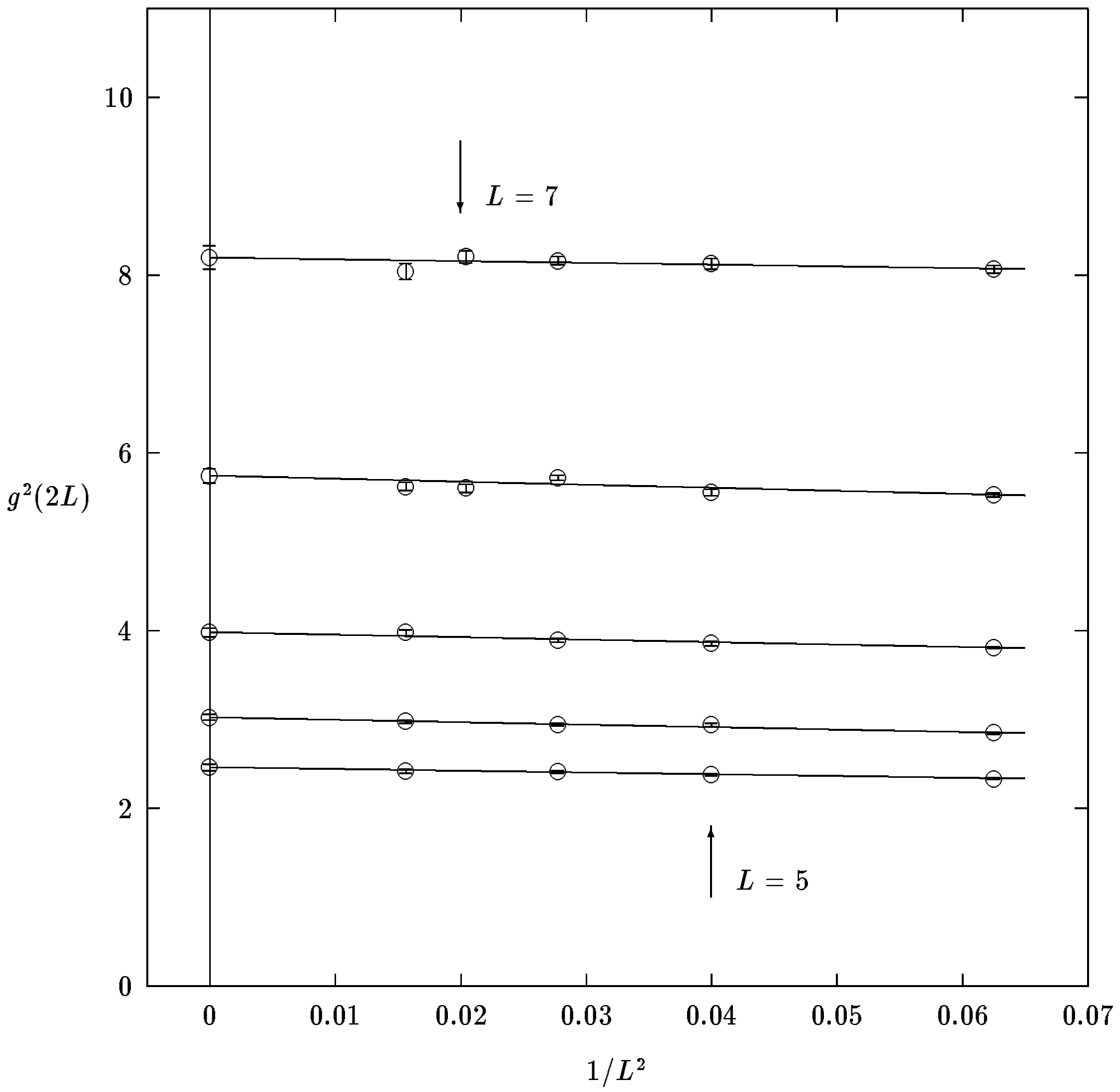

Figure 1: The extrapolation to the continuum limit

| $g^2(L_{phys})$ | $g^2(2L_{phys})$ (numer.) | $g^2(2L_{phys})$ (2–loops) |
|---|---|---|
| 2.075 | 2.459(38) | 2.428 |
| 2.430 | 3.024(31) | 2.938 |
| 3.068 | 3.980(52) | 3.955 |
| 4.108 | 5.741(76) | 6.001 |
| 5.597 | 8.20(13) | 10.48 |

Table 3: The values for $g^2(2L_{phys})$ compared with the two–loops ones

Table 3 shows the numerical values extrapolated to the continuum limit of $g^2(2L_{phys})$ and a comparison with the values predicted by the two–loop evolution for that step. The continuum limit is obtained by making a least–squares fit of the data for $g^2(2La)$ with a curve linear in $L^{-2}$.

Figure 2 shows the values of the renormalized coupling extrapolated to the continuum limit as a function of an energy scale in arbitrary units.

Besides the error on the extrapolation to the continuum limit of table 3, there is a second source of errors when the different values of the renormalized couplings of table 3 are put in sequence to reconstruct the running coupling constant. Indeed, each recursive step has an error on the coupling at the large scale which must be propagated through the various steps. This is done by converting the uncertainty on the coupling $g^2(2L_{phys})$ into an uncertainty on the renormalization scale which then would fluctuate around an exact multiple of two. However, the renormalized coupling at the scale $L_{phys}$ of the step $n+1$ is not exactly equal to the one at the scale $2L_{phys}$ of the step $n$. By using the two–loops expression for the $\beta$–function, this small difference can be turned into a slight modification of the renormalization scale which will fluctuate at the various steps around a value different from an exact multiple of two.

This is reported in table 4, by taking as the reference value for the scale the one of the fifth iteration step, where we will perform the calibration of the scale in physical units. In figures 2 and 4, the errors on the scale are turned again into errors on the couplings. These figures contain points at scale lengths which are without error while a cumulative error bar appears on the values of the coupling. The dashed line in figure 2 is the behaviour predicted by the two–loops continuum $\beta$–function normalized to the coupling at highest energy which closely follows the running coupling up to the last points and justifies the use of the two–loops $\beta$–function in the error propagation procedure described above.

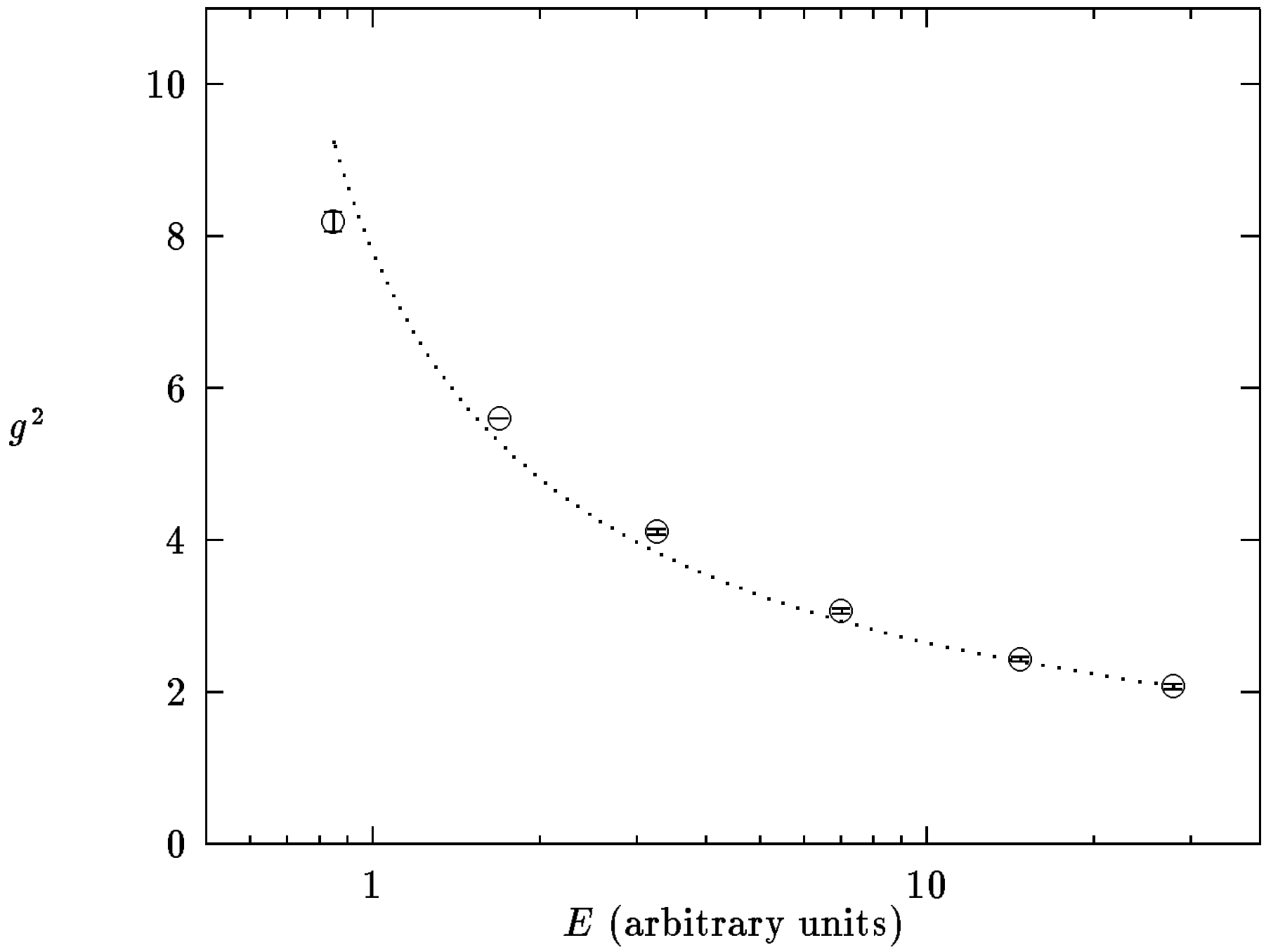

Figure 2: The running coupling constant as a function of the energy scale in arbitrary units

| $L_{phys}/L_0$ | $g^2(L_{phys})$ | $E \equiv 1/L_{phys}$ (Gev) |
|---|---|---|
| 0.061(5) | 2.075 | 27.9(2.3) |
| 0.115(6) | 2.430 | 14.8(8) |
| 0.242(9) | 3.068 | 7.02(26) |
| 0.520(11) | 4.108 | 3.27(7) |
| 1.000 | 5.597 | 1.70 |
| 2.000(32) | 8.200 | 0.85(1) |

Table 4: The running coupling constant as a function of the scale given in units of the reference scale $L_0$ and in GeV

The calibration of the length scale, *i.e.* of the lattice spacing, requires a precise calculation of a physical quantity on the largest volume and close enough to the continuum limit. In our case we have not performed such a dedicated calculation which is certainly crucial for the real unquenched SU(3) case. Following ref. [3] we have taken the available values of the string tension at a $\beta$ value 2.85 overlapping with the values of $\beta$ of the fifth step of the recursive procedure. This can be seen from figure 3 which shows the dependence of $\beta$ upon $L$ at fixed renormalized coupling: the line shows a fit with the dependence implied by asymptotic scaling which is in very good agreement with the results of the simulation. The calibration should be taken as an exercise rather than as a precise determination: indeed such a $\beta$ value lies between the calculations of the renormalized coupling on a $L = 4$ and an $L = 5$ lattice. The mild $L$ dependence of our results for the renormalized coupling justifies the utility of such an exercise. A careful normalization will be discussed elsewhere [8]. By adopting for the string tension a conventional value of (425 MeV)$^2$ we get that the scale for the running coupling constant shown in figure 4 and in the third column of table 4 is in GeV. The lower points in figure 4 correspond to the calculation performed in ref. [3] with the same scale calibration. The two definitions should match in the perturbative region, once after implementing the relative correction term. This can be included by rescaling the calibration of one scale with respect to the other by the ratio of the $\Lambda$ parameters given by:

$$\Lambda_{TP}/\Lambda_{SF} = 3.6206(2) \quad , \tag{8}$$

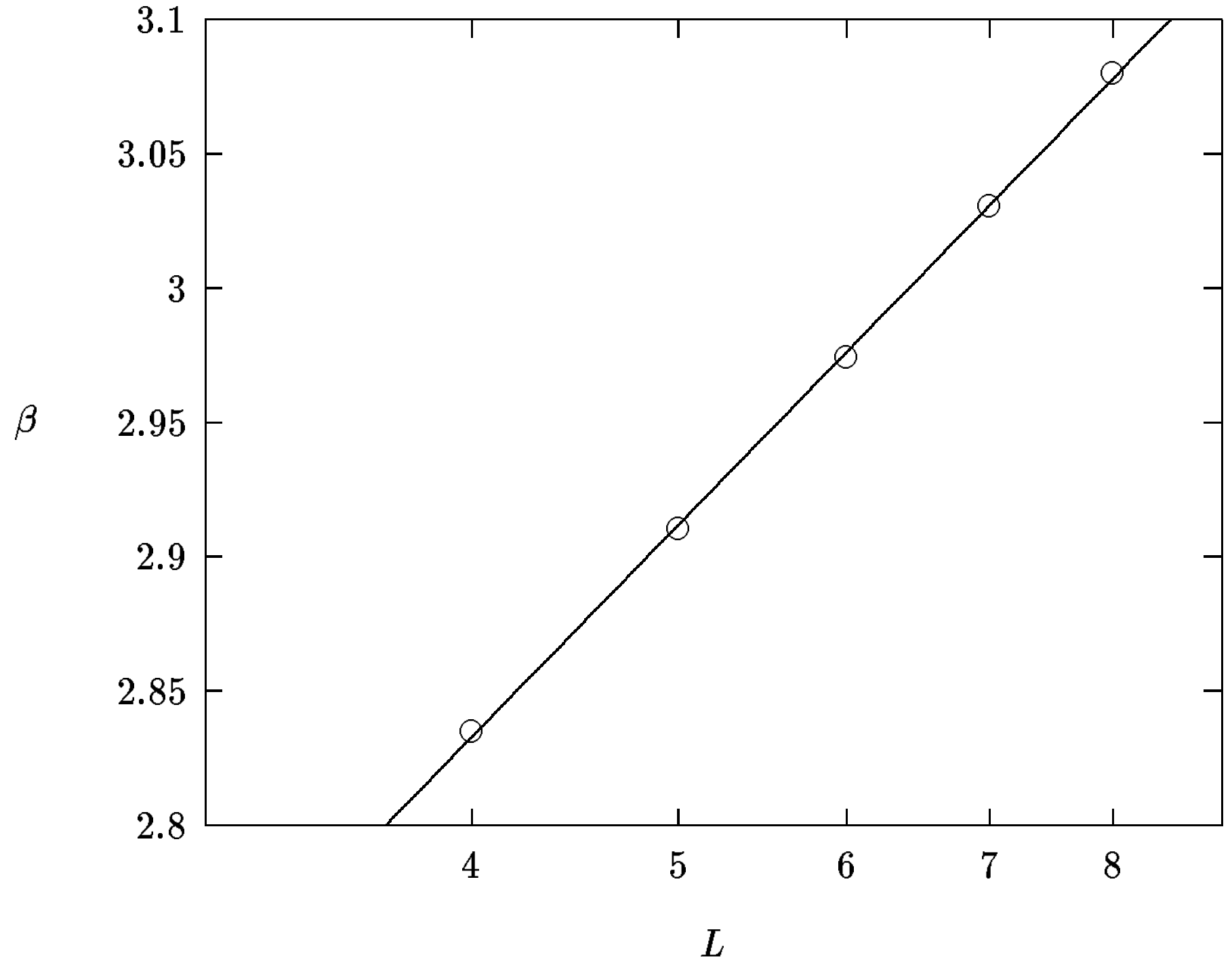

Figure 3: The dependence of $\beta$ upon the number of points $L$ at constant renormalized coupling $g^2(L_{phys}) = 5.597$

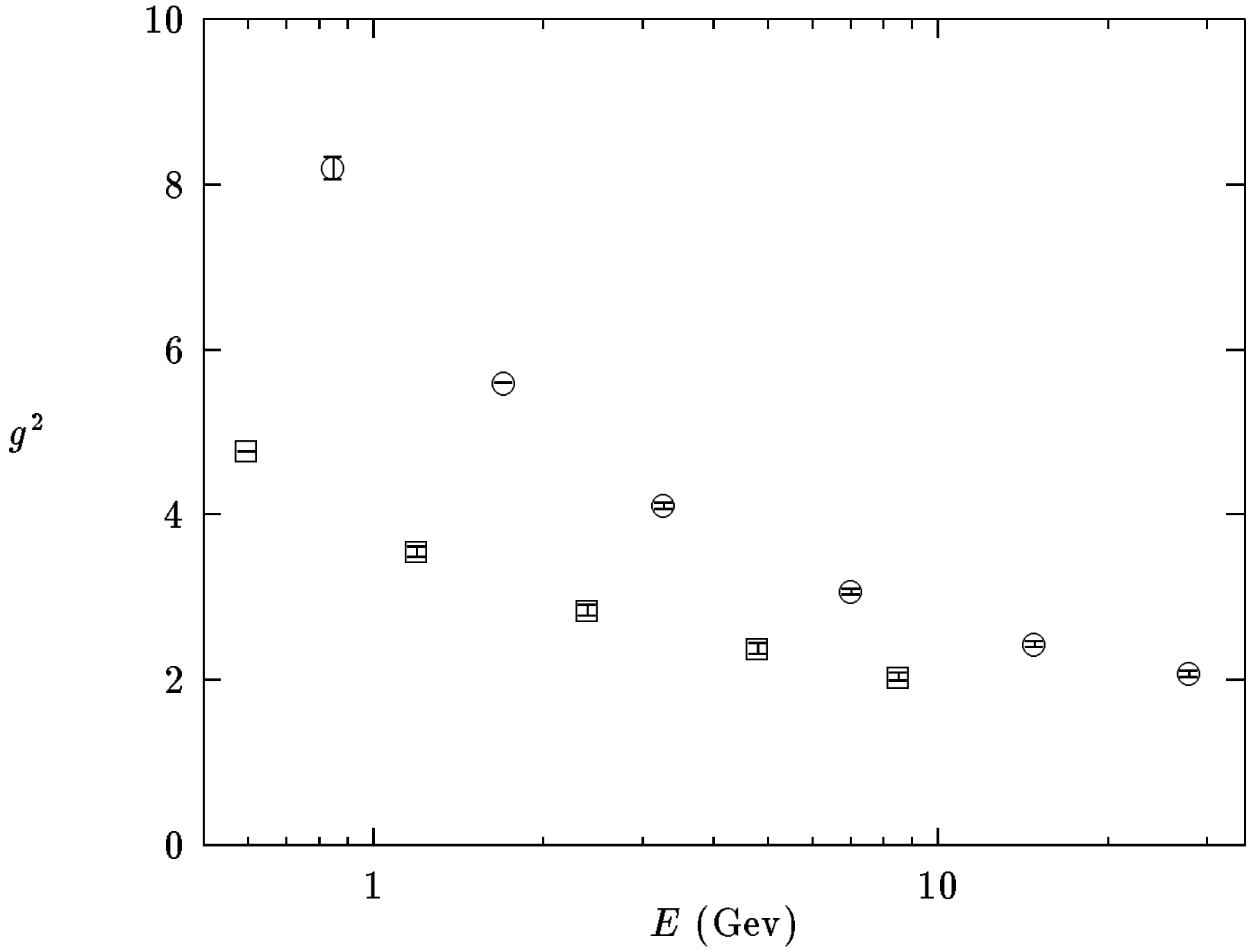

Figure 4: The running coupling constants defined in this paper (○) and the one of ref. [3] (□) as a function of the energy scale with the same lattice calibration

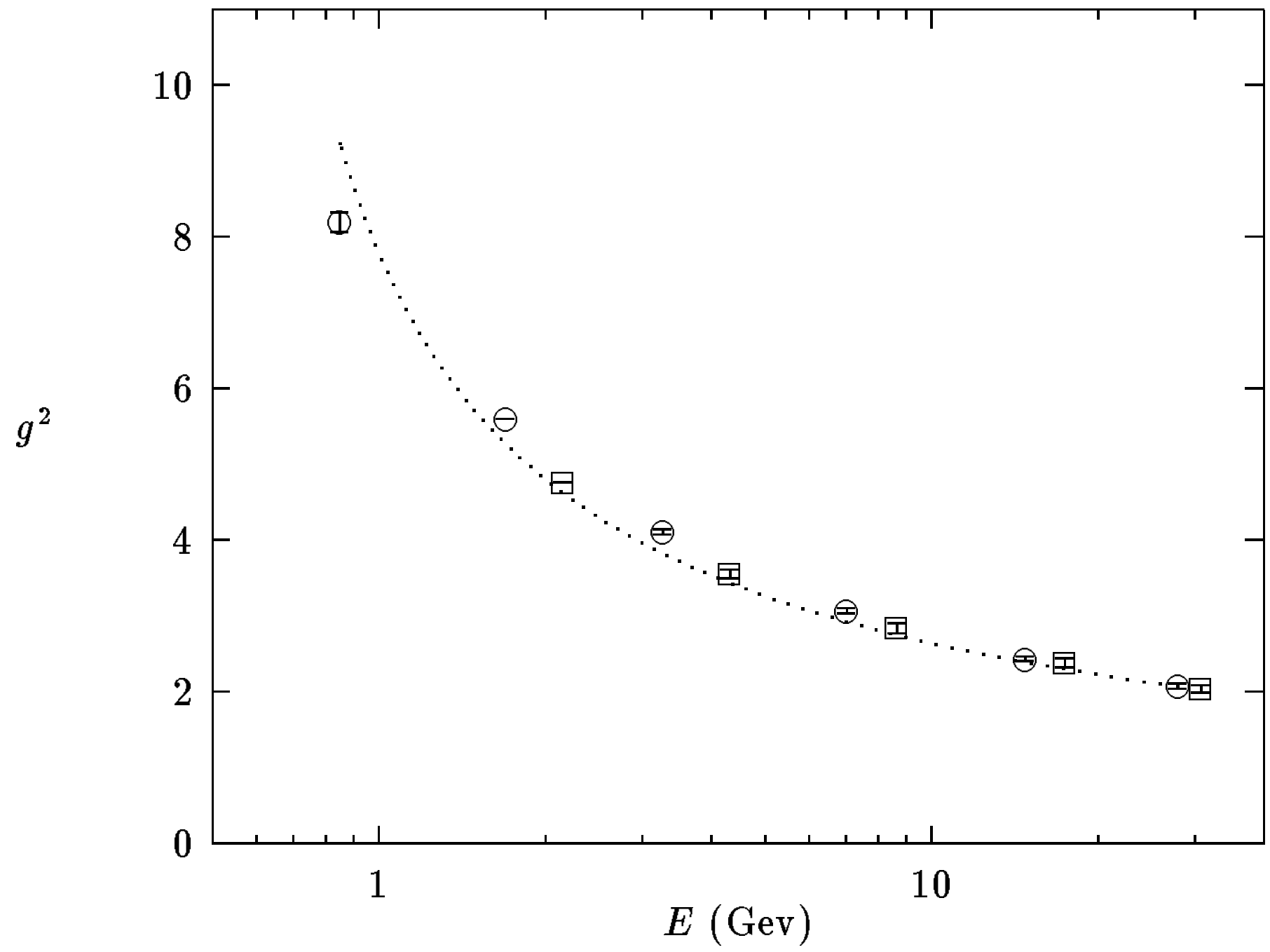

Figure 5: The running coupling constants defined in this paper (○) and the one of ref. [3] (□) with the same lattice calibration after rescaling the energy scale by the ratio of $\Lambda$ parameters

where $\Lambda_{SF}$ is the $\Lambda$ parameter of the definition based on the Schrödinger functional (see ref. [2]).

Figure 5 shows the result of this rescaling: at small values of the coupling the two definitions nicely superimpose to each other and, partly surprisingly, they do so also at rather large values of the coupling.

A result independent from $\Lambda$ is the $\beta$–function: by locally fitting the running coupling constant with a polynomial one can extract the $\beta$–function for our definition and the one of ref. [3]. As an exercise we fitted three points at time with a second–order polynomial in $\log(E)$: the result is plotted in figure 6, together with the two–loops $\beta$–function. Beyond a value of the

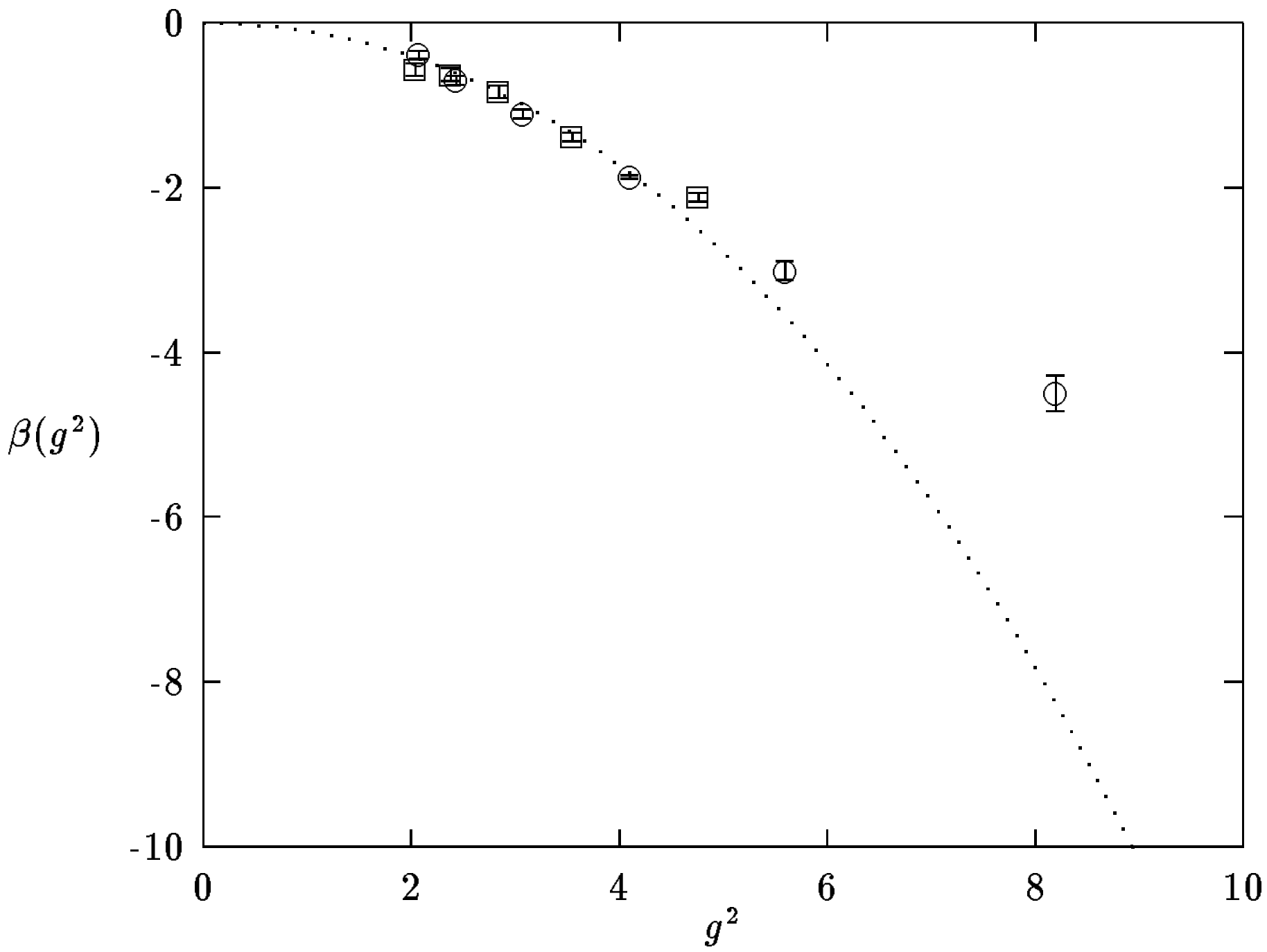

Figure 6: $\beta$–function for the running coupling constants defined in this paper ($\circ$) and the one of ref. [3] ($\Box$) is compared with the two–loops expression (dotted line)

coupling around 6 our $\beta$–function deviates from the two–loops result.

# 5 Conclusions

The above simulation shows the feasibility of a calculation of the running strong coupling constant at a few percent level. The calculation took about 1,600 CPU hours at an average speed of about two Gigaflops on a three Gigaflops version of the APE100 series. The unquenched case at least requires, with the present algorithms, a factor hundred in computer time. Moreover, it requires the understanding of the inclusion of fermions with twisted boundary conditions for which interesting suggestions have been made.

## Acknowledgements

The authors thank for interesting and stimulating discussions M.Lüscher, R.Sommer, U.Wolff and P.Weisz. The calculation could not be performed without the constant support of the APE group at the University "La Sapienza" and in particular of S. Cabasino, F. Marzano, P. S. Paolucci, R. Sarno, M. Torelli, G. M. Todesco and P. Vicini.

## References

[1] M. Lüscher, P. Weisz and U. Wolff, Nucl. Phys. **B359** (1991) 221.

[2] M. Lüscher, R. Narayanan, P. Weisz and U. Wolff, Nucl. Phys. **B384** (1992) 168.

[3] M. Lüscher, R. Sommer, P. Weisz and U. Wolff, Nucl. Phys. **B389** (1993) 247.

[4] M. Lüscher, R. Sommer, P. Weisz and U. Wolff, Nucl. Phys. **B413** (1994) 481.

[5] G. 't Hooft, Nucl. Phys. **B153** (1979) 141.
A. Gonzalez–Arroyo,M. Okawa, Phys. Rev. **D27** (1983) 2397.
P. van Baal, Twisted Boundary Conditions: ..., Thesis, Utrecht (July 1984);
P. van Baal and A. S. Kronfeld, Nucl. Phys. **B 9** (Proc. Suppl.) (1989) 227.

[6] M. Lüscher and P. Weisz, Nucl. Phys. **B266** (1986) 309.

[7] G.M. de Divitiis, R. Frezzotti, M. Guagnelli and R. Petronzio, ROM2F-93-43 preprint, to be published in Nucl. Phys. B.

[8] The ALPHA–Collaboration, in preparation.

[9] K. Fabricius and O. Haan Phys. Lett. **143B** (1984) 459.

[10] A.D. Kennedy and B.J. Pendleton Phys. Lett. **156B** (1985) 393.